\documentclass[a4paper]{elsarticle}

\title{A Mathematically Robust Model of Exotic Pine Invasions}

\author[1]{Elliott Hughes}

\author[1]{Miguel Moyers-Gonzalez}
\author[1]{Rua Murray}

\author[1,2]{Phillip L.\! Wilson\corref{cor1}}
\cortext[cor1]{Corresponding author, email: phillip.wilson@canterbury.ac.nz}

\affiliation[1]{organization={School of Mathematics and Statistics, University of Canterbury},
addressline={20 Kirkwood Avenue, Upper Riccarton},
city={Christchurch, 8041},
country={New Zealand}}

\affiliation[2]{organization ={Te Punaha Matatini}, 
city={Auckland}, 
country= {New Zealand}}

\usepackage{amsmath}
\usepackage{amssymb}
\usepackage{graphicx}
\usepackage{commath}
\usepackage{textcomp}
\usepackage{gensymb}
\usepackage{float}
\usepackage[colorlinks=true,linkcolor=blue,citecolor=blue]{hyperref}
\usepackage{caption}
\usepackage{subcaption}
\usepackage{natbib}
\usepackage{tikz}
\usetikzlibrary{arrows,automata}
\usetikzlibrary{positioning}
\usetikzlibrary{arrows.meta,positioning}
\usetikzlibrary{patterns}
\usetikzlibrary{patterns.meta}
\usepgflibrary[patterns]
\usepackage{pgfplots}
\usepgfplotslibrary{fillbetween}
\pgfplotsset{width=0.25\textwidth,compat=1.9}

\usepackage{chngcntr}
\counterwithin{equation}{section}
\usepackage[nameinlink,capitalize]{cleveref}
\usepackage{makecell}

\begin{document}

\begin{abstract}
  Invasive pine trees pose a threat to biodiversity in a variety of Southern Hemisphere countries, but understanding of the dynamics of invasions and the factors that retard or accelerate 
  spread is limited. Here, we consider the past models of wilding pine spread and develop a new model of pine invasion. We show that many prior models feature parameter estimates 
  which are not biologically supported and rely on a conjecture to obtain an asymptotic spread speed of invasive pine populations, the main output of these models. In contrast to 
  prior approaches, we use partial differential 
  equations to model an invasion. We show that invasions are almost static for a significant period of time before rapidly accelerating to spread at a 
  constant rate, matching observed behaviour in at least some field sites. 
  Our work suggests that prior methods for estimating invasion speeds may not accurately predict spread and are sensitive to assumptions about the distribution of 
  parameters. However, we present alternative estimation methods and suggest directions for further research.
\end{abstract}
\maketitle

\section{Introduction}
Many pine varietals are popular forestry species across the globe for their fast growth rates and long, straight trunks \cite{richardson2006pines}. 
Unfortunately, because of these desirable properties, some pine species also pose significant threats to native environments \cite{richardson1994pine,richardson1998ecology,MPI_pine_report}. 
Significant invasions have been reported in New Zealand, Australia, South Africa, Chile, and other countries \cite{taylor2016drivers}. Because pine species are 
frequently both economically important for forestry and major invasive threats, management strategies often focus on containing invasions and regulating 
new pine plantings to limit invasive risks \cite{richardson2006pines} (see \cite{MPI_pine_report} for an example). For management purposes, understanding 
the invasive potential of different plantings is particularly important. Consequently, prior literature modelling exotic pine invasions has often focused on 
estimating the spread rates of invasions in different environments (i.e.\! the rate at which the boundary of an invaded region moves in a particular direction over time). 

\par 
Several papers have produced mathematical models for exotic conifer invasions over the last twenty years, employing a variety of different approaches 
and focusing on different research questions. The goal of this paper is to review this literature, particularly focusing on models that produce estimates 
for spread rates, before presenting a new approach to modelling pine invasions. This new partial differential equation (PDE) model has several advantages, including 
allowing for explicit simulation of invasions and a greater variety of analytical tools compared to past integrodifference matrix (IDM) or cellular automa models.

\subsection{Determinants of Wilding Pine Spread}
In many cases, past literature exploring invasions of wilding pines (e.g.\! the colonisations of landscapes by pines) focuses primarily on the spread rate of 
such invasions (that is, the rate at which the boundary of a colonised region moves outward). 
The spread rates of pines are determined by a combination of species-specific and environmental factors \cite{pauchard2016pine, richardson1998ecology}, since pine seeds are wind-dispersed and the spread rate of trees depends on terrain and climatic conditions \cite{richardson1994pine}. 
For example, seeds from trees on top of a hill 
will generally be dispersed much further than those in a valley below \cite{richardson1994pine}, increasing the initial spread rates of invasions beginning on elevated 
terrain. Many 
pine species release seed during specific times of the year, so weather conditions 
during these periods are an important determinant of spread \cite{coutts2012reproductive}. Some pine species are serotinous, with cones that only open to release 
seeds during hot temperatures (e.g.\! greater than $45\degree$C) \cite{agee1998fire} but solar heating can lead to cone temperatures in this range, 
even when external air temperatures are more temperate \cite{wyse2019seed}. For example, in New Zealand field conditions  
50\% of cones released seed even in sites where temperatures never exceeded $30\degree$C \cite{wyse2019seed}, it 
is likely that this feature merely constrains seed release to periods where average temperatures are high \cite{wyse2019seed}. In non-serotinous species seeds are also dispersed during a relatively narrow window \cite{keeley1998evolution} 
(in New Zealand seed dispersal occurred between July and September in studied locations).

\par
Most pines also rely on partner fungal species which limits spread where these fungi are not present \cite{dickie2010co}. Local variation in site suitability 
is an important factor determining spread rates, although there is some evidence these effects are usually dominated by those stemming from landscape-scale weather effects \cite{pauchard2016pine}. 
Pre-existing vegetation is also an important determinant of spread, with substantial invasions usually limited to grassland or shrubland areas \cite{ledgard2002spread}.

\par
Compared to other trees with wind-borne seeds, pines have several features that make them particularly prone to spread. Many pine species reach reproductive 
maturity at unusually young ages \cite{richardson2006pines}, and can generate large amounts of seed per season (e.g.\! over 10,000 seeds per year for a fully-mature tree) \cite{coutts2012reproductive}. 
Species with highly favourable characteristics have been known to disperse seeds over very long distances (up to 25km in one extreme case) \cite{richardson1994pine}. Observed spread rates have exceeded 100m/year, so invasions can occur very quickly under 
optimal conditions \cite{buckley2005slowing}. Population growth rates are thought to exhibit negative density dependence, so invading populations 
will increase most rapidly when their numbers are relatively small \cite{sprague2021density}. 

\subsection{Management of Wilding Pines}
Pine trees are very long lived once established \cite{keeley1998evolution} and it is extremely 
labour intensive to eliminate entrenched populations \cite{nunez2017ecology}. Effective management strategies seek to identify high risk areas and invading 
populations early in their development \cite{nunez2017ecology}. 
A variety of such management strategies for pine trees have been proposed (see for example \cite{moody1988controlling} or \cite{maxwell2009rationale}). Some approaches 
attempt to reduce spread by eliminating `source' populations of established trees, as these 
contribute more seed than peripheral populations \cite{caplat2014cross}. Other strategies focus on eliminating these peripheral populations on 
the theory that trees further from the source spread seed further than those in source populations \cite{caplat2014cross}.

\par
Since these management strategies frequently 
require mutually exclusive areas of focus, it is necessary to prioritise some proposed strategies over others \cite{caplat2014cross}. To determine which strategies produce 
better results, data is needed either on the performance of these differing strategies or to parametrise simulations that might enable 
comparisons to be made in a simulated environment. Existing models for spread rates neglect important components of spread, including 
all environmental factors and any within population variation in seed production. Consequently, more research is needed to enhance understanding of these basic questions and to 
provide predictions of spread to policymakers.

\section{Mathematical Models of Wilding Pine Spread}
Existing models for wilding pine spread 
differ considerably depending on their intended purpose. Indeed, given the range of ecological questions modellers may wish to answer and the diversity of ecological systems, it is hardly surprising that in general a wide range of 
mathematical tools have applications to modelling the spatial spread of invasive species. Frameworks range from graph-theoretic 
sandpile models \cite{caldarelli1998modelling}, to cellular automata \cite{caplat2014cross}, 
to differential equations on lattices \cite{messinger2013predator}, to partial differential equations 
\cite{holmes1994partial}. For a discussion of some of the most common models, the interested 
reader should consult \cite{thompson2021mechanistic}.

\par
At the most abstract level, there are many papers attempting to use statistical (or machine learning) models to estimate the 
effect of various parameters on pine establishment, growth rates or other outcomes (e.g.\! \cite{sprague2021density, pauchard2016pine}). Such models are useful as they enable inferences to be made 
about the significance of different parameters on spread, but they do not attempt to produce completely accurate models of invasion dynamics. Other models do attempt, to some degree, to capture these underlying dynamics 
(e.g.\! \cite{buckley2005slowing, caplat2012seed, caplat2014cross}). Within the wilding pine  
literature a wide variety of techniques and approaches are used, although there are some notable 
omissions.

\par
In particular (to the best of the authors' knowledge) all prior mathematical models of wilding pine spread are discrete in time. Such discrete models can be justified when dispersal happens at a particular stage in an individual organism's life cycle 
and individuals are otherwise relatively sedentary \cite{KOT1986109} (note, however, that this does not imply they are inappropriate in other cases). Some such models (e.g.\! \cite{caplat2014cross}) also discretise in space, using cellular automata to 
simulate spread, while others employ continuous spatial variables (such as \cite{buckley2005slowing} and \cite{caplat2012seed}). Whether or not to discretise 
in space largely depends on the intended function of the model. Spatial discretisation enables relatively straightforward analysis 
of situations where management strategies or other factors vary spatially (by allowing easy simulation of different spatial configurations, for example), while a continuous space framework 
eliminates the artificial boundaries caused by spatial discretisation. Furthermore, depending on the 
exact structure of the model, continuous or discrete approaches also allow for a range of different 
mathematical tools to be employed (e.g.\! graph theory or calculus for some discrete or continuous approaches).

\par
While discrete time models are certainly straightforward to simulate, the rich theory of continuous-time 
models has been comparatively neglected. In particular, while there is a large literature on 
PDE models for spatial spread, such models are infrequently fitted to empirical data or applied to specific problems \cite{holmes1994partial}. 
Often the only class of PDE discussed for modelling spatial spread are 
reaction diffusion equations, which represent only a subset of the universe of PDE 
models \cite{neubert2000demography}. Our intention in the latter part of this paper is to demonstrate that 
more general PDE models can provide useful insights.

\par
Beginning a more detailed analysis of the prior literature \cite{caplat2014cross}, uses a spatially discrete approach to simulate the 
response of pine invasions to different management strategies (e.g.\! those that target peripheral stands first versus those 
that begin in areas of highest density). The model presented in \cite{caplat2014cross} divides a hypothetical region into a grid of sub-regions, each of which 
is assumed to contain some non-negative number of individual pine trees (which in turn each belong to cohorts according to their 
age). Beginning from a small number of subregions populated with adult trees, the population in each region evolves 
according to a set of deterministic rules of growth and dispersal. Management strategies can be simulated by choosing to 
eliminate all trees from specific regions at each time-step, where target regions are selected by according to some rule approximating the 
management strategy in question. Habitat fitness in each region can also be straightforwardly simulated, a 
key advantage of this approach over other wilding pine models (e.g.\! \cite{buckley2005slowing} or \cite{caplat2012seed}). It is important to note, however, 
that parameters in this model were not chosen to match observed parameters in actual wilding pine invasions, so the results of 
\cite{caplat2014cross} should not be considered as predictive of rates of pine spread under 
different management strategies or habitat suitabilities.

\par
\citeauthor{davis2019simulation} later extended this work to model the 
response of \emph{P.\! contorta} to wildfires in Patagonia \cite{davis2019simulation}. In both New Zealand and Patagonia, 
\emph{P.\! contorta} planted for erosion control or timber purposes now pose a threat to native 
temperate grasslands \cite{richardson1994pine}. While \citeauthor{davis2019simulation} do not attempt to 
produce estimates of spread rates, their work does consider otherwise neglected factors like spatial herterogeneity in 
site suitability. In particular \citeauthor{davis2019simulation} modify the model of \cite{caplat2014cross} 
to include multiple vegetation types and fire events. Since Patagonian steppe includes both 
grasslands and patches of beech (\emph{Nothofagus antarctica}), cells in the model are 
randomly allocated to either vegetation type \cite{davis2019simulation}. Germination of wilding 
pines is reduced by 90\% in cells populated by \emph{N.\! antarctica} but this effect is reduced 
in cells which have recently experienced a fire \cite{davis2019simulation}. 

\par
\citeauthor{davis2019simulation} allow 
one fire to occur during each simulation, with a fire beginning in a random cell and spreading 
probabalistically to surrounding cells \cite{davis2019simulation}. The fire destroys all pine trees in that cell but 
increases the probability of pine trees invading that cell in following years, as other types 
of vegetation are assumed to be knocked back by fire events \cite{davis2019simulation}. \citeauthor{davis2019simulation} 
found that these fires significantly increased spread in the Patagonian environment, although this 
relied upon a sufficient prior density of pine trees before the fire occurred \cite{davis2019simulation}. 
The impact of wildfires is likely much lower in some other contexts such as New Zealand, where a much smaller area 
is affected each year on average \cite{meyer2021winds}. However there is some evidence that pine invasions 
may increase the risk of wildfires \cite{taylor2017pinus} and climate change may also increase 
the risk of such events \cite{meyer2021winds}. However, because the model of \cite{davis2019simulation} is parametrised for 
the Chilean context and is intended to estimate spread when wildfires occur, it does not provide appropriate estimates of 
spread rates in other contexts. Furthermore, \citeauthor{davis2019simulation} do not account for negative density dependence, instead assuming that pine growth 
rates in a cell are independent of the number of other trees present (at least when the population is less than the carrying capacity).

\par
Two other papers (\cite{buckley2005slowing} and \cite{caplat2012seed}) attempt 
to predict spread rates of wilding pines under New Zealand conditions. Both of these papers use the IDM (integrodifference matrix model) framework of \citeauthor{neubert2000demography} (as presented in \cite{neubert2000demography}) 
to produce a model with an explicit population structure and separated growth and dispersal components. It is important to note, however, that neither of these 
papers attempt to produce a nonlinear model for spread, but instead proceed directly to construct a linear model for spread on the assumption 
that this linear model is an appropriate linearisation of an underlying nonlinear model. Consequently both works 
rely upon the results of \cite{weinberger1982long} that, under certain conditions, the asymptotic spread speeds of species in a nonlinear integrodifference model 
are equal to the asymptotic spread speeds in that model's linearisation. Furthermore, these models also rely upon the conjecture of \citeauthor{neubert2000demography} 
that the maximum spread rates computed in \cite{neubert2000demography} are also the asymptotic spread rates of these linearised models. 

\par 
Taking these assumptions as given, both papers simulate spread with the following integrodifference matrix equation 

\begin{equation}\label{eq:linear_model_diffusion}
    n_{t+1}(x) = \int_{-\infty}^{\infty}\left(K(x - y) \circ A\right)n_t(y)dy
\end{equation}

\noindent 
Where $n_t(x)$ is a vector of abundances such that the $i$-th element of this vector is the population in demographic class $i$ at location $x$ at time $t$, $a_{ij}$ is the $ij$-th element of $A$ giving the proportion of 
organisims in class $i$ that transition to class $j$ at the end of one year, $k_{ij}$ is the dispersal kernel associated with this state transition 
and $\circ$ denotes element-wise multiplication. For further 
discussion the reader is encouraged to consult \cite{neubert2000demography}, \cite{weinberger1982long} or \cite{mythesis}.

\par 
The behaviour of this model also depends critically on the dispersal kernels selected for the reproducing categories. Selection of dispersal kernels is in 
itself a complex topic and the reader is encouraged to consult \cite{nathan2011mechanistic} or another suitable text for an overview. Setting these general 
issues aside, however, we will proceed now to consider the specific models utilizing this IDM approach for pine invasions.

\section{The Model of Buckley et al.}\label{sec:buckley_model}
Prior to the more recent work of \citeauthor{caplat2012seed}, the first paper employing an IDM model for wilding pines was that of \cite{buckley2005slowing}. While \cite{buckley2005slowing} considers models for pine spread in ungrazed grassland, grazed grassland, and 
shrubland, we will limit our discussion to the case of ungrazed grassland for ease of comparison with \cite{caplat2012seed}. The demographic structure of this model 
has eight classes, although five of these have essentially identical roles. There are six juvenile classes and organisms in these classes do not reproduce 
and have a constant survival probability $s_J$. At the end of each year an organism in this class advances to the next class, except in the final class where there is a constant retention rate 
$q$. An advantage of separating out juveniles into these six separate classes is that it sets a minimum speed at which organisms can become reproductively mature, so that an organism must be at 
least six years old before it begins to reproduce. There are two reproducing classes which produce $f_i$ viable seeds (with $f_1 < f_2$ so younger organisms produce less seeds) that have a constant 
fixing probability $b$. Organisms are retained in the first class at a rate $r$ and there is a constant survival probability $s_A$ for organisms in either of these classes. Consequently the linear 
population projection operator is 

\begin{equation}
    A = 
    \begin{bmatrix}
        0 & 0 & 0 & 0 & 0 & 0 & f_1b & f_2b \\
        s_J & 0 & 0 & 0 & 0 & 0 & 0 & 0 \\
        0 & s_J & 0 & 0 & 0 & 0 & 0 & 0 \\
        0 & 0 & s_J & 0 & 0 & 0 & 0 & 0 \\
        0 & 0 & 0 & s_J & 0 & 0 & 0 & 0 \\
        0 & 0 & 0 & 0 & s_J & s_Jq & 0 & 0 \\
        0 & 0 & 0 & 0 & 0 & s_J(1-q) & s_Ar & 0 \\
        0 & 0 & 0 & 0 & 0 & 0 & s_A(1-r) & s_A 
    \end{bmatrix}.
\end{equation}

\begin{figure}[H]
	\centering
	\includegraphics{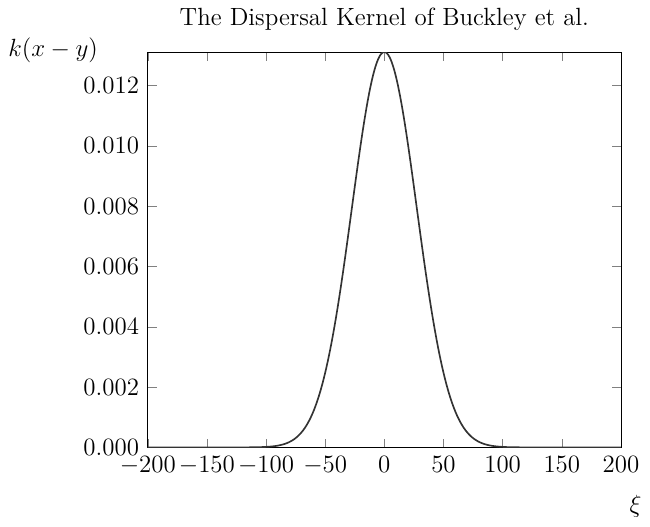}
	\caption{The dispersal kernel used in \cite{buckley2005slowing}, with the parameters set to the point estimates given in \cite{buckley2005slowing} (e.g.\! $\sigma = 27.39$, $\beta = 1287$, $p = 0.1$).}
	\label{fig:buckley_kernel}
\end{figure}

\noindent The corresponding matrix of dispersal kernels for each stage transition is 

\begin{equation}
    K(\xi) = 
    \begin{bmatrix}
        0 & 0 & 0 & 0 & 0 & 0 & k(\xi) & k(\xi) \\
        \delta(\xi) & 0 & 0 & 0 & 0 & 0 & 0 & 0 \\
        0 & \delta(\xi) & 0 & 0 & 0 & 0 & 0 & 0 \\
        0 & 0 & \delta(\xi) & 0 & 0 & 0 & 0 & 0 \\
        0 & 0 & 0 & \delta(\xi) & 0 & 0 & 0 & 0 \\
        0 & 0 & 0 & 0 & \delta(\xi) & \delta(\xi) & 0 & 0 \\
        0 & 0 & 0 & 0 & 0 & \delta(\xi) & \delta(\xi) & 0 \\
        0 & 0 & 0 & 0 & 0 & 0 & \delta(\xi) & \delta(\xi) 
    \end{bmatrix},
\end{equation}

\noindent where $\xi = x-y$ and $k(\xi)$ is given as a mix of normal and negative exponential distributions so that $k(\xi) = (1-p)k_{\text{short}}(\xi) + pk_{\text{long}}(\xi)$. Given that one of these 
distributions is defined on the entire real line and the other is defined only on the positive part of the real line, the resulting 
distribution must be the following piecewise expression

\begin{equation}
    k(\xi) = \begin{cases}
        (1-p)k_{\text{short}}(\xi), & \xi < 0, \\
        (1-p)k_{\text{short}}(\xi) + pk_{\text{long}}(\xi), & \xi \geq 0,
    \end{cases}
\end{equation}

\noindent
where $k_{\text{short}}$ is a normal distribution with mean zero and standard deviation $\sigma$ and $k_{\text{long}}$ 
is a negative exponential distribution with shape parameter $\beta$. A seed enters the long-distance distribution with 
probability $p$ and otherwise enters the short-distance distribution. It should be noted 
that this imposes the assumption that long-distance spread can only occur in one direction, presumably that corresponding to the predominant wind direction. 
It also assumes that seed 
dispersed from young trees travels as far as that from older trees, unlike \cite{caplat2012seed}. These long and short-distance distributions 
are phenomenological, rather than following the mechanistic WALD 
model used in \cite{caplat2012seed}. See \autoref{fig:buckley_kernel} 
for a sketch of the dispersal kernel for the point estimates given in \cite{buckley2005slowing} 
(noting in particular the almost symmetrical nature of the distribution).

\par
As in later papers, this model is parametrised in units of seeds (or trees), metres, and years. Consequently one should interpret 
all demographic parameters (e.g.\! those in $A$) as rates per year and all dispersal 
parameters ($\sigma$, $\beta$, $p$) as determining the spread of seeds in metres from the source. Since the dispersal kernels used 
in \cite{buckley2005slowing} are continuous distributions, each iteration of the model deposits fractional 
amounts of seed at different locations which obviously does not reflect biological reality. The most reasonable interpretation of 
these fractional seed counts at different locations is to understand them as reflecting averages or expectations of seed quantities 
in each location, although it is not clear that this framework will converge to the same spread rates as would a fully stochastic simulation. 
Setting aside such concerns, it is useful to briefly consider how the spread rates are estimated in \cite{buckley2005slowing}.

\par
To obtain a point estimate of spread rates under uncertainty in parameter values, \cite{buckley2005slowing} samples 10,000 different 
combinations of parameters and calculates the spread rate (following the approach of \cite{neubert2000demography}) for each combination. From these 10,000 
different spread rates, \citeauthor{buckley2005slowing} take the median spread rate as a point estimate and obtain $95\%$ confidence intervals by sampling from the resulting distribution of 
spread rates. This approach requires 
the estimation of distributions for each of these parameters. 

\subsection{Parameter Estimation}
Since \cite{buckley2005slowing} had low levels of data for most parameters, in general parameters are calculated by 
taking a triangular distribution with a median point estimate matching that of the observed data and maximum and minimum values drawn from those observed in the 
field \cite{buckley2005slowing}. 
In some cases in which more data was available distributions are apparently calculated based on best fit to the observed data, although it is not specified which parameter distributions are calculated 
in this manner \cite{buckley2005slowing}. Demographic parameters are estimated based on prior fieldwork and, in some cases, existing literature where data could not be obtained 
\cite{buckley2005slowing}. 

\par
Tree fecundity is estimated by calculating the number of cones produced by each tree, then estimating the number of viable seed contained in each cone. Trees are divided into two 
classes according to age, although it is unclear how this segmentation is done. It should be noted that this approach assumes that all seeds produced by the tree are spread along some straight line (which is the same for all trees in the 
landscape), 
which seems unlikely in practice even in locations where winds almost always blow in particular direction. Heterogeneity in seed production both between individual trees and across time is ignored, although it is 
stated that tree sampling likely occurred during an unusually productive year \cite{buckley2005slowing}. Seedling survival is estimated based on a field experiment where seeds were sown in plots and 
survival was regularly assessed \cite{buckley2005slowing}. Data on minimum and maximum ages of reproductive maturity is estimated from field observations and an assumption is made about the 
median value to obtain a rough distribution for $q$. Adult survival rates for pine trees are based on overseas data, as no data for New Zealand were available at the time of writing \cite{buckley2005slowing}. Dispersal 
parameters for the long and short-distance distributions are based on field work and on analysis of two aerial photos respectively \cite{buckley2005slowing}. As is hopefully clear, the very low 
levels of data available for parametrisation meant that robust estimates of parameters and their distributions are not obtained \cite{buckley2005slowing}.

% \begin{landscape}% Landscape page
\begin{table}[H]
\begin{center}
\begin{tabular}{|c|c|c|c|c|}
    \hline
    Name & Description & Source \\
    \hline
    Demographic parameters & & \\
    \hline
    $s_A$ & Adult survival & Prior literature \\
    $f_1$ & Sub-Adult seed production & Field data \\
    $f_2$ & Adult seed production &  Field data \\
    $b$ & Seed establishment & Field data \\
    $q$ & Retention in juvenile classes & Field data \\
    $r$ & Sub-Adult retention & \makecell{Not parametrised \\ from data} \\ 
    $s_J$ & Juvenile survival & Field data \\
    \hline 
    Spread parameters & & \\
    \hline
    $p$ & \makecell{Probability of \\ long-distance dispersal} & Prior literature \\
    $\beta$ & \makecell{Exponential distribution \\ shape parameter} & Aerial photography \\
    $\sigma$ & \makecell{Normal distribution \\ shape parameter} & Aerial photography \\
    \hline
\end{tabular}
\caption{Parameter descriptions and sources from \cite{buckley2005slowing}.}
\end{center}
\end{table}
% \end{landscape}

\par
Ultimately, \citeauthor{buckley2005slowing} estimate a median 1,500m/year spread rate and a $95\%$ confidence interval of (1,190, 1,957) \cite{buckley2005slowing}. It should be noted 
that this spread rate seems to overestimate observed spread in New Zealand settings, including from the site which was used to parametrise spread rates (the interested reader should consider 
Figure 1 in \cite{buckley2005slowing} and compare the spread from between 1965 to 1980 with the mean spread rate from the model). \citeauthor{buckley2005slowing} also attempt to compute the impact of changes in various parameters 
on observed spread rates \cite{buckley2005slowing}, but given the high levels of inaccuracy in the predicted spread rate it is not clear that this sensitivity analysis 
will produce accurate results.

\section{The Model of Caplat et al.}\label{sec:caplat_model}
We now move to analysing the particular system developed in \cite{caplat2012seed}, the most recent attempt to model pine invasions using the model of 
\citeauthor{neubert2000demography}. 
In this model, the population is divided into four classes; seedling, juvenile, sub-adult and adult, rather than the eight in \cite{buckley2005slowing}. 
The resulting population projection matrix is given by 

\begin{equation}
    A = \begin{bmatrix}
        0 & 0 & e_sf_1 & e_sf_2 \\
        s_J & s_Jr_J & 0 & 0 \\
        0 & s_J(1-r_J) & 0 & 0 \\
        0 & 0 & s_A & s_A 
    \end{bmatrix},
\end{equation}

\noindent where $e_s$ is a constant establishment probability of seeds, of which sub-adults produce $f_1$ and 
adults produce $f_2$ each year. These seedlings have a probability $s_J$ of surviving to become juveniles, 
which they do at the end of one year. Juveniles also have the same probability of surviving $s_J$ 
and are retained in the juvenile class at a rate $r_J$. Sub-adults have a survival probability $s_A$ 
and become adults at the end of one year, while adults have the same constant survival probability $s_A$. Note 
that this matrix implies that a small proportion of seedlings begin seed production in two years, 
which is not biologically supported \cite{buckley2005slowing, coutts2012reproductive, richardson1998ecology}. 
\par
The dispersal matrix for this system is 

\begin{equation}
    K(x-y) = 
    \begin{bmatrix}
        0 & 0 & k_1(x-y) & k_2(x-y) \\
        \delta(x-y) & \delta(x-y) & 0 & 0 \\
        0 & \delta(x-y) & 0 & 0 \\
        0 & 0 & \delta(x-y) & \delta(x-y)
    \end{bmatrix},
\end{equation}

\noindent where $k_1$ and $k_2$ are Wald distributions, parametrised following the method of \cite{katul2005mechanistic}. In particular for $x > y$ the dispersion 
kernel is given by 

\begin{equation}
    k_i(x-y) = \sqrt{\frac{\gamma_i}{2\pi(x-y)^3}}\exp\left(-\frac{\gamma_i(x-y-\mu_i)^2}{2(x-y)\mu_i^2}\right),
\end{equation}

\noindent where $\mu_i$ and $\gamma_i$ are parameters calculated as follows 

\begin{equation}
    \mu = \frac{\overline{U}h_i}{v_s}, \qquad \gamma_i = \frac{\overline{U}h_i^2}{2\kappa h_c\sigma}.
\end{equation}

\noindent Here $\overline{U}$ is the mean windspeed and $\sigma$ is the standard deviation of windspeed, $h_1$ is the seed release height for sub-adult trees, $h_2$ is the seed release height for adult trees, $h_c$ is the canopy height, 
and $\kappa$ is a turbulence coefficient that varies between 0.3 and 1. 
It should be noted that these 
distributions are defined only for strictly positive values of $x-y$. This leads to the strong assumption that spread will only occur in one direction, which is unlikely to 
be fulfilled in practice. Furthermore, since fecundity is calculated in an identical way to \cite{buckley2005slowing}, this leads to the assumption that all seeds 
from each tree are dispersed only in one direction. 

\begin{figure}[H]
	\centering
	\includegraphics{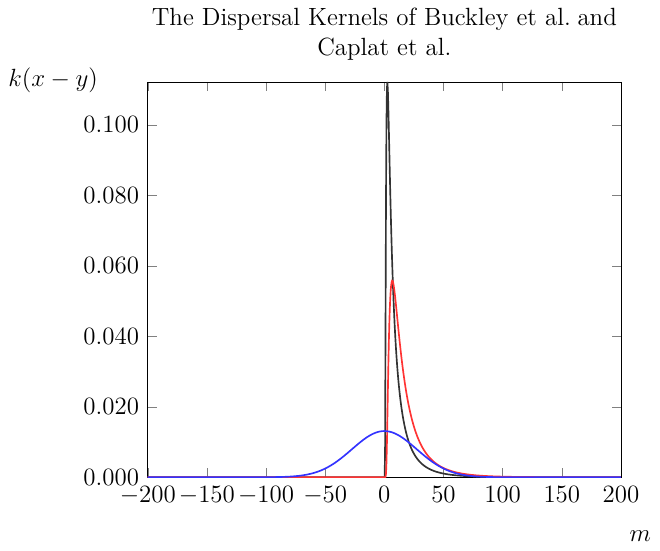}
	\caption{Comparison of the dispersal kernel of \cite{buckley2005slowing} (blue), 
    with the sub-adult dispersal kernel of \cite{caplat2012seed} (red) and the adult dispersal 
    kernel (black).}
	\label{fig:caplat_kernel}
\end{figure}

\subsection{Parameter Estimation}
Parameter distributions are taken from \cite{buckley2005slowing} in the case of adult survival and establishment 
probabilities or otherwise obtained from field data gathered 
from one location between 2008 and 2009. Wind data was taken from two locations within the field area as the Northern part of the site was more sheltered by the landscape than the Southern part.

%\begin{landscape}% Landscape page
    \begin{table}[H]
    \centering
    \makebox[\textwidth]{
    \begin{tabular}{|c|c|c|c|}
        \hline
        Name & Description & Source \\
        \hline
        Demographic parameters & & \\
        \hline
        $s_A$ & Adult survival & \citeauthor{buckley2005slowing} \\
        $f_1$ & Sub-Adult seed production & Field data \\
        $f_2$ & Adult seed production &  Field data \\
        $r_J$ & Seed establishment & \citeauthor{buckley2005slowing} \\
        $e_s$ & Retention in juvenile classes & Field data \\
        $s_J$ & Juvenile survival & Field data \\
        \hline 
        Spread parameters & & \\
        \hline
        $h_c$ & Canopy height & Field data \\
        $h_{t1}$ & Seed release height for sub-adults & Field data \\
        $h_{t2}$ & Seed release height for adults & Field data \\
        $\kappa$ & Synthetic parameter & \makecell{Not parameterised \\ from data} \\
        $v_s$ & Terminal velocity of seeds & Experimental data \\
        $U$ & Mean windspeed & Field data \\
        $\sigma_w$ & Standard deviation of windspeed & Field data \\
        \hline
    \end{tabular}
    }
    \caption{Parameter descriptions and sources from \cite{caplat2012seed}.}
    \end{table}
%\end{landscape}    

\subsection{Issues with Prior Parameter Estimates}
It is unclear why some parameter estimates are selected. For example \cite{caplat2012seed} states that the retention parameter for the juvenile class 
is calculated such that at the estimated age of first reproduction $2\%$ of surviving trees remained unproductive \cite{caplat2012seed}. Assuming that this estimated age of first reproduction is intended to be a 
reasonable estimate of the average age of first reproduction this choice seems difficult to defend. In practice, if one considers the central estimate given for this parameter in \cite{caplat2012seed} this 
suggests that in a mature forest (that is, in a case where the age of trees is random), the age of first reproduction for adult trees in this forest is distributed according to a geometric distribution with $p = 0.22$. 
Thus in a forest of mature trees the modal age of first reproduction is two years, which is not biologically supported \cite{buckley2005slowing, coutts2012reproductive, richardson1998ecology}. 
It should also be noted that canopy height is presumably a function of population density in practice, so 
for a linearisation around 0 (as is the case for \cite{neubert2000demography}) the canopy height parameter should be very near zero. Thus, given 
the parameter choices for canopy height are explicitly taken between the 80-90th percentile it is not clear that \autoref{eq:linear_model_diffusion} is a 
valid linearisation for the appropriate underlying nonlinear model with the parameter distributions 
given in \cite{caplat2012seed}.

\par
Considering briefly the point estimates given in Table 1 of \cite{caplat2012seed} 
it is easy to see that these produce strange results. In particular, an adult tree is assumed to produce 
approximately 2,000 viable seeds of which approximately $90\%$ will fall along some straight line 
that extends 50 metres in one direction from the tree. Consequently about $1,800\times0.96 = 1,728$ seedlings will survive to the juvenile stage 
one year after the initial seed fall from the adult tree. In the following year $1,728\times0.2112 \approx 364$ 
juveniles will advance to sub-adulthood (most of the rest will be retained in the juvenile category). 
These sub-adults will also produce seeds and in the following year there will be approximately 355 
adults within 50 metres of the adult tree. If these trees were distributed uniformly this would imply a density of approximately 1 adult tree every 
30cm within three years, which seems unreasonable. Consequently one is left to assume that the underlying nonlinear model 
must feature strongly density dependent growth, although this does not resolve all possible concerns (e.g.\! the assumption that 
1800 viable seeds will fall within 50m of a source tree, all on a single line). 

\par 
Because of these parameter issues and the inherent instability of such linear models, actual 
simulations of this model tend to produce uncontrolled exponential growth (although obviously 
this depends on the exact parameter values and initial condition chosen). As an example, we 
consider a model where each parameter is set at the central estimate given in \cite{caplat2012seed}. 
Since this model is defined in terms 
of discrete counts of trees, a natural initial condition would be zero everywhere except for on 
some sets of measure zero where it would take on positive integer values. Of course, from a 
practical perspective this implies no dispersal and so is clearly unsuitable. Furthermore, 
since adult pine trees might be expected to be metres or more in diameter, they necessarily take 
up more than a particular location on the real line. 

\par
\begin{figure}[H]
    \centering
    \includegraphics{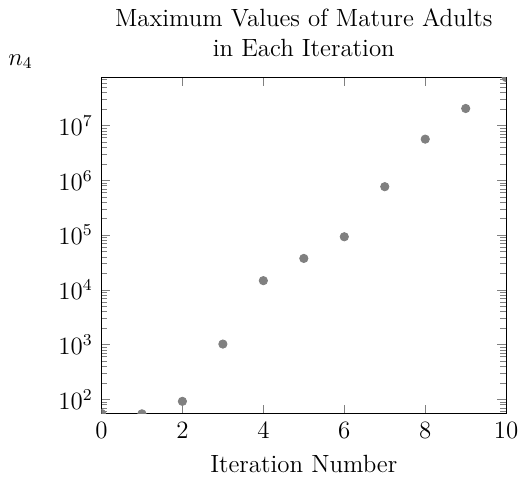}
    \caption{$\max_{x \in \mathbb{R}}n_4(x)$ for the first ten iterations  
    on a logarithmic scale.} 
    \label{fig:simulate_caplat}
\end{figure}

 
Therefore, we will return to an idea discussed in \autoref{sec:buckley_model}; that the value of $n_i$ at some 
particular location is an average or expectation over some portion of the landscape. For our 
initial condition we will consider an expression of the form 

\begin{equation}
    n(x,0) = \begin{bmatrix}
        0 \\
        0 \\
        0 \\
        T(x)
    \end{bmatrix},
\end{equation}

\noindent 
where $T(x)$ is a triangular distribution from $x = -2$ to $x = 2$ with a mean of zero. Without 
being very precise, we can consider this as a spatial averaging of a single adult tree over a 
small region. Furthermore it is a continuous function with a compact support of positive measure 
and so it is in the class of initial conditions considered in \cite{neubert2000demography}. Furthermore, 
with an appropriate spatial discretisation of the domain, we can approximate the behaviour of 
the linear system computationally. For simplicity, integrals were approximated with the trapezium 
rule and a spacing of $\Delta_x = 10^{-1}$. With this approximation to the action of the linear 
operator we can compare the maxima of the approximate solutions for adult trees across the 
first ten iterations (see 
\autoref{fig:simulate_caplat}).

\noindent 
As we can see, this linear model produces unreasonable results with maximum values becoming 
unphysically large after three years. Because this model experiences approximately exponential growth (at 
least with this initial condition and parameter values) simulations with this linear model are not 
good approximations of the true behaviour of the invasive process. We also tested that this exponential growth occurs with other initial 
conditions (e.g. hat functions, positive constant initial conditions and point masses, as well as varying the demographic classes of initial 
organisms) confirming that this observed behaviour is not an artifact of our particular choice of condition. Because of this consistent 
pattern of explosive growth these models are forced to rely 
on the asymptotic spread rate calculation of \citeauthor{neubert2000demography} (see \cite{neubert2000demography} or \cite{mythesis}). 

\subsection{Results}
Notwithstanding these issues, \citeauthor{caplat2012seed} follow \cite{buckley2005slowing} and compute 
spread rates by sampling 10,000 different parameter combinations and calculating a mean spread rate across all of these parameter combinations. This analysis 
is repeated for four possible wind directions (NE, NW, SE and SW) and is found to be in reasonable agreement with data from the field location \cite{caplat2012seed}. However, these estimates 
are more accurate for the Northern part of the site than for the the Southern part (where estimated spread rates are approximately $60\%$ of those actually observed from 1965 onwards, when spread began in this part of the site). \citeauthor{caplat2012seed} 
suggest that this discrepancy is largely due to higher interannual variation in wind statistics, which will not be captured in their data \cite{caplat2012seed}. It is nonetheless worth 
noting that in the Northern part of the site the terrain was not flat and so the WALD model was not well suited to describing 
seed dispersal within this area \cite{caplat2012seed}.

\par
To make clear the implications of the possible inaccuracies 
we consider an alternative specification where every parameter is distributed uniformly. Where 
maximum and minimum values were reported these are taken as the bounds of the uniform distribution, 
while if a standard deviation was reported this was taken as the standard deviation of a uniform distribution 
centred around the reported point value and from this maximum and minimum values were inferred. Under these assumptions the mean spread rate was 370m/year, which is 
several times the mean spread rate reported by \cite{caplat2012seed}. 

\par
The overall implication of the proceeding section is that the results of \cite{caplat2012seed} 
are extremely sensitive to a variety of assumptions, many of which are difficult to defend. 
Consequently it is important to be cautious when considering the management implications suggested in \cite{caplat2012seed}. In particular, 
\cite{caplat2012seed} presents elasticities of spread rates with respect to each of the 
parameters in their model but these elasticities naturally depend on current parameter 
values. Thus, incorrect simulated distributions for parameters or incorrect transition 
matrices may bias the importance of some parameters over others. 

\section{A New Model for Wilding Pine Spread}
Given the limitations of previous models outlined above, we now introduce a new 
model for wilding pine spread. We believe our approach has several advantages over previous models 
as it allows for explicit simulations of invasions and is easily extensible to handle previously 
neglected factors. While PDE are sometimes neglected by ecologists, 
the richer mathematical theory and existing computational tools that are available for these 
equations simplify analysis compared to the IDM approach used in previous models. In this section, we present the new model and conduct perturbation analysis to obtain a semi-analytic approximation of the solution.

\par
Unlike previous models, we pose our model in terms of the density of trees in particular 
demographic categories at particular locations. One advantage of this approach is that it ensures that the variables 
of interest are continuous, rather than discrete quantities or counts as in previous models (e.g.\! \cite{buckley2005slowing, caplat2012seed}). 
This clarifies interpretation of fractional quantities of interest (see \autoref{sec:buckley_model}) and ensures that we can appropriately apply a 
PDE model to this scenario.

\par
In this density formulation, we consider two categories or subdivisions. In particular, we divide the quantity of biomass in a particular region into 
reproducing biomass (which we call as a short hand adult biomass or $A$) and potential reproducing biomass (as a shorthand child biomass or $C$). Note that here we are not saying that biomass 
in the adult category is dedicated to reproduction (e.g.\! comprised entirely of cones, seeds, etc) but rather that it comes from a fully mature 
tree. Meanwhile, child biomass should primarily be thought of as a measure of the quantity of just mature biomass that will appear at a particular location as the 
present immature trees at that location mature. We adopt this slightly unwieldy convention because we intend to let this variable 
flow freely around the landscape, rather than attempting to track the location and life-stage of particular non-reproducing trees. Thus treating $C$ as a measure of `potential' 
biomass is the most appropriate approach, although this does necessarily complicate the question of scaling (see below). 

\par
For scaling purposes it is also useful to be precise about the quantity we are measuring when we refer to the density of biomass in a particular location. Due 
to greater availability of data \cite{madgwick1994pinus} we consider the biomass in our model to refer to above ground dry biomass. This choice simplifies 
parameter estimation and scaling arguments, but necessarily imposes some assumptions. In particular, this approach forces us to assume that root structure 
development is closely coupled to all observable tree outcomes, so that it is sufficient to track above ground biomass in order to predict future behaviour 
of the system. By tracking dry biomass, rather than the `wet' biomass as observed \emph{in situ}, we are also making a similar assumption with respect to water 
content.

\par
Given the very long lifetime of adult pine trees in New Zealand we assume that once pine trees are established at some location, that 
location will tend towards some equilibrium level of biomass. In particular, in the absence of children, we assume that the growth rates 
of adult biomass are approximately quadratic (with parameters $\rho_0,\kappa > 0$) so that growth rates are initially positive before becoming negative beyond some critical biomass level. 
It is well known that the differential equation 

\begin{equation}
    \frac{dA}{dt} = \rho_0A - \kappa A^2,
\end{equation}

\noindent produces a S-shaped or sigmoidal trajectory for density over time (given a small but non-zero initial density of trees). Thus one 
consequence of this assumption is that the growth rate of a small population of adult trees will initially accelerate, before approaching zero 
as the population approaches the carrying capacity of the environment. We will also assume 
that child biomass transitions to the adult category at some fixed rate $\mu$, so that the growth rate of adult biomass increases linearly with 
increasing amounts of child biomass.

\par
Children are produced in areas containing adult trees at a density dependent rate that 
is bounded above by some maximum production rate. Dispersal from source locations is 
diffusive and children leave this state at a fixed rate. This leads to the following equation for $C$; 

\begin{equation}
    \frac{\partial C}{\partial t} = D\frac{\partial^2 C}{\partial x^2} - \mu C + g(A),
\end{equation}

\noindent where $g(A)$ is a production function for adult biomass. 
While allowing `child' biomass to disperse may seem rather unphysical, since we are viewing $C$ as a 
measure of potential biomass in a particular location we can feel more confidence in allowing this quantity to move around the landscape with time. Furthermore, since child biomass neither produces seeds nor 
has density dependent effects in our model, there is no particular need to know the exact location of non-adult trees. The alternative, that child biomass is fixed and some `seed' category of mobile biomass is introduced, was considered but such a 
category would necessarily feature dynamics that evolve on a timescale of minutes or hours, as opposed to the multi-year processes that drive outcomes among the 
`adult' category. Consequently this approach was rejected as computationally and analytically infeasible.

\subsection{The Implied Dispersal Kernel}
One implication of this 
diffusive dispersal and fixed transition rate is that, in the absence of secondary 
growth, a single burst of seed deposited in a specific location (on an infinite domain) will be 
dispersed such that the final density distribution will follow a Laplace kernel. Note this final density distribution is  
analogous to the dispersal kernel considered in both \cite{buckley2005slowing} and \cite{caplat2012seed} (see \cite{lewis2016mathematics}). To see this we will follow 
the approach of \cite{lewis2016mathematics} and take advantage of the fact that adult biomass does not move over time so that we can use a simplified version of 
our full model to obtain a dispersal kernel. In particular, we will ignore 
any deaths that occur in the juvenile state, secondary seed production, and growth that occurs after biomass has 
transitioned to the adult state ($\rho_0 = \kappa = 0$). For simplicity we will restrict our analysis to a 1D model, although the calculation is 
essentially the same in two dimensions. This leads to the very simple dimensional model 

\begin{align}\label{eq:disp_kernel_deriv_1}
    \frac{\partial A}{\partial t} &= \mu C, \\
    \frac{\partial C}{\partial t} &= -\mu C + D\frac{\partial ^2 C}{\partial x^2}, \label{eq:disp_kernel_deriv_2}
\end{align}

\noindent the initial condition $C(x,0) = \delta(x)$, $A(x,0) = 0$ creates the desired point mass and is coupled with an infinite domain (under the usual condition that solutions are exponentially damped). Finally, because we wish for all seed 
to transition to the adult state, we also impose that $C(x,\infty) = 0$. Our dispersal kernel will now simply be the asymptotic distribution of $A$ as $t \rightarrow \infty$. 
If one integrates the right-hand side of the previous equation for $C$ this leads to the following expression 

\begin{equation}
    \int_0^{\infty}\frac{\partial C}{\partial t}dt = \left[C\right]_0^{\infty} = \lim_{t \rightarrow \infty} C(x,t) - C(x,0) = 0 - \delta(x).
\end{equation}

\noindent Given this, integrating both sides of \Cref{eq:disp_kernel_deriv_1,eq:disp_kernel_deriv_2} gives 

\begin{align}
    A(x,\infty) &= \int_0^{\infty}\mu Cdt, \\
    -\delta(x) &= \int_0^{\infty}-\mu C + D\frac{\partial ^2 C}{\partial x^2}dt.
\end{align}

\noindent Letting $k(x) = A(x,\infty)$ we can substitute this expression into the second equation to obtain 

\begin{equation}
    -\delta(x) = -k + \frac{D}{\mu}\frac{\partial^2 k}{\partial x^2}.
\end{equation}

\noindent This equation is a simple linear ODE, which can be solved by Fourier transforms. This leads to the following expression for the dispersal kernel  

\begin{equation}
    k(x) = \frac{a}{2}e^{-a|x|},
\end{equation}

\noindent often known as a Laplace distribution. Note this dispersal kernel implies that spread follows a process which is isotropic (there is no preferred direction of spread) 
and that dispersal is leptokurtic (tails are heavier than if the dispersal kernel was normally distributed). Furthermore, the 
leptokurticity of the dispersal kernels of this model overcomes one of the main objections often advanced about classical PDE models for spread (that 
dispersal is assumed to be Gaussian).

\par
Returning to our particular model, we can see that combining all the interactions we have discussed so far leads to the following coupled system of differential equations

\begin{align}
    \frac{\partial A}{\partial t} &= \mu C + f(A), \\
    \frac{\partial C}{\partial t} &= D\nabla^2C - \mu C + g(A),
\end{align}

\noindent with the following parameters 

\begin{table}[H]
    \centering
\begin{tabular}{|c|c|c|c|c|}
    \hline
    Name & Description \\
    \hline
    $\mu$ & Promotion rate from children to adults \\
    $f(A)$ & Adult survival and rejuvenation \\
    $D$ & Diffusion of children \\
    $g(A)$ & Adult reproduction \\
    \hline
\end{tabular}
\caption{Parameters in the general model.}
\end{table}

\begin{figure}[H]
    \centering
    \includegraphics{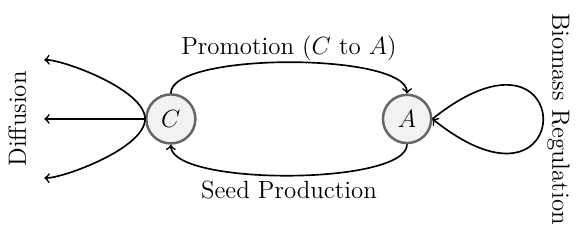}
    \caption{Conceptual diagram of the interactions forming our model.}
    \label{fig:concep_diag}
\end{figure}

\noindent Conceptually, we represent this model with the following diagram (\autoref{fig:concep_diag}),

\par
In practice we will make several simplifications. Most importantly we will assume $g(A) = \gamma A^2/(\beta^2 + A^2)$ 
for some parameters ($\gamma$ and $\beta$). This form of $g$ implies that 
for small $A$ growth rates in $C$ are approximately linear in $A^2$, but that as $A$ increases the 
growth rate of $g(A)$ decreases and total production will never exceed $\gamma$. Note here we will treat $\gamma$ as the 
maximum rate of viable potential biomass production (we will not track any biomass that will not survive to adulthood). 
As we have already discussed above, we will set $f(A) = \rho_0 A - \kappa A^2$ so that growth follows a sigmoidal 
trajectory.

\par
Finally, we will consider the 
case with only one spatial dimension, leading to the following equations 

\begin{align}
    \frac{\partial A}{\partial t} &= \mu C + \rho_0A - \kappa A^2, \\
    \frac{\partial C}{\partial t} &= D\frac{\partial^2 C}{\partial x^2} - \mu C + \frac{\gamma A^2}{\beta^2 + A^2}.
\end{align}

\par
With this system in hand, we now need to consider appropriate boundary and initial conditions 
for our model. 
In this paper we shall restrict ourselves to considering spread in a valley or similar 
landscape, where spread is restricted to occur within some closed region $[-L,L]$. This implies the no-flux boundary 
conditions

\begin{equation}
    C_x(t,-L) = C_x(t,L) = 0.
\end{equation}

Moving now to the initial conditions, we shall consider the case where spread occurs from a 
shelterbelt or small stand of at least partially mature trees. We will assume that in this band of older trees, biomass 
is close to the equilibrium level and non-zero outside. This leads to the initial condition 

\begin{equation}
    A(x,0) = A_0h_n(x) = \begin{cases}
        \frac{\rho_0}{\kappa} & x \in [-L/n,L/n], \\
        0 & \text{otherwise},
    \end{cases}
\end{equation}

\noindent where $n \in \mathbb{R}^+$. Note that by imposing this initial condition we are assuming that at equilibrium the contribution of biomass 
from new trees maturing is negligible compared to the total amount of equilibrium biomass. However, in the following section we will see that this assumption is reasonable.

In this case we expect 
$C(0,x) = 0$, except possibly at the origin (but for simplicity we will take $C(0,x) = 0$). Thus our combined initial conditions are

\begin{align}
    A(0,x) &= A_0h_n(x), \\
    C(0,x) &= 0.
\end{align}

\subsection{Nondimensionalisation}
We will assume that there exists some uniform (spatially invariant) equilibrium (time invariant) solution $(A_*,C_*)$ in the positive quadrant (this will be shown to hold later, 
at least given suitable assumptions on the order of parameters). We chose this uniform equilibrium state as the starting point for my nondimensionalisation on the theory 
that solutions to the model with biologically realistic initial conditions should generally tend towards this state as $t \rightarrow \infty$. This reflects the 
fact that we expect pine populations to invade any surrounding grassland, leading asymptotically to a uniformly forested landscape. In order for $C_*$ to be time-invariant we can require that 
$\mu C_* \sim \frac{\gamma A^2_*}{\beta^2 + A_*^2}$. For $A_*$ to be time-invariant, we can 
further require that either $\rho_0A_* \sim \kappa A^2_*$ and $\mu C_* \ll \rho_0A_*$ or $\mu C_* \sim \kappa A^2_*$ and $\rho_0 A_* \ll \kappa A^2_*$ or 
$\mu C_* \sim \rho_0 A_* \sim \kappa A^2_*$.

\par
Disposing of a simple case first, we can see that if $A \ll \beta$ then 
then production of child biomass at equilibrium approaches zero while if $\beta \ll A$ then seed production approaches a switching function 

\begin{equation}
    f(A) = \begin{cases}
        0 & A = 0, \\
        \gamma & A \neq 0.
    \end{cases}
\end{equation}

\noindent But in practice it is known that young trees tend to be less productive than older trees, so we would expect total biomass to affect seed generation across the 
range of plausible values for $A$ and  
thus that $\beta \sim A_*$. A natural consequence of this and our non-simplification assumption is that $\gamma \sim \mu C_*$. Given these assumptions and based on data from New Zealand and abroad, 
estimates for the magnitude (in terms of years $T$, weight per square meter $kg/m^2 = M$ and meters $L$) of each parameter can be obtained (see \autoref{tab:param_magnitudes}). 

\begin{table}[H]
    \centering
\begin{tabular}{|c|c|c|c|c|}
    \hline
    Name & Scale & Units \\
    \hline 
    $\rho_0$ & $10^{-1}$ & $T^{-1}$ \\
    $\kappa$ & $10^{-3}$ &  $M^{-1}T^{-1}$ \\
    $\mu$ & $10^{-1}$ & $T^{-1}$ \\
    $\gamma$ & $10^{-1}$ & $MT^{-1}$\\ 
    $\beta$ & $10^2$ & $M$ \\
    $A_*$ & $10^2$ & $M$ \\
    $C_*$ & $1$ & $M$ \\
    \hline
\end{tabular}
\caption{Scales of parameters and equilibrium states.}
\label{tab:param_magnitudes}
\end{table}

\noindent For a derivation of these estimates please see \cite{mythesis}. 

\subsection{Scaling}
Proceeding now to nondimensionalise the system we let $\eta_A \tilde{A} = A$, $\eta_C \tilde{C} = C$, $\eta_t \tilde{t} = t$ and $\eta_x \tilde{x} = x$ 
to obtain the system 

\begin{align}
    \frac{\partial \tilde{A}}{\partial \tilde{t}} &= \frac{\mu\eta_t\eta_C}{\eta_A} \tilde{C} + \rho_0\eta_t \tilde{A} - \kappa\eta_t\eta_A \tilde{A}^2, \\
    \frac{\partial \tilde{C}}{\partial \tilde{t}} &= \frac{D\eta_t}{\eta_x^2}\frac{\partial^2 \tilde{C}}{\partial \tilde{x}^2} - \mu \eta_t \tilde{C} + \frac{\eta_t}{\eta_C}\frac{\gamma \tilde{A}^2}{\beta^2/\eta_A^2 + \tilde{A}^2}.
\end{align}

Since we are interested primarily in dispersal, we desire that our final scaling has $A_*$, $C_*$ are $\mathcal{O}(1)$ and that $\mu C_* \sim \gamma A_*^2/(\beta^2 + A_*^2) \sim 1$. To work out some desirable scales, consider the uniform equilibrium 
states of these coupled equations. These are solutions to the equations

\begin{align}
    0 &= \frac{\mu\eta_t\eta_C}{\eta_A} \tilde{C}_* + \rho_0\eta_t \tilde{A}_* - \kappa\eta_t\eta_A \tilde{A}_*^2, \\
    0 &= - \mu \eta_t \tilde{C}_* + \frac{\eta_t}{\eta_C}\frac{\gamma \tilde{A}_*^2}{\beta^2/\eta_A^2 + \tilde{A}_*^2}.
\end{align}

\noindent Given the difference in magnitude of $A_*$ and $C_*$ we expect that a suitable scaling will result in a negligible $\mu C$ term at equilibrium and thus

\begin{equation}
    \rho_0 \tilde{A}_* - \kappa \eta_A \tilde{A}_*^2 \approx 0.
\end{equation}

\noindent Thus for $\tilde{A}_* \sim 1$ we set that $\eta_A = \rho_0/\kappa \sim 10^2$ (which matches our previous assumptions). Solving for $\tilde{C}_*$ leads to 

\begin{equation}\label{eq:tilde_C_eqn}
    \tilde{C}_* = \frac{\gamma}{\mu\eta_C}\frac{\tilde{A}_*^2}{\beta^2/\eta_A^2 + \tilde{A}_*^2},
\end{equation}

\noindent Given our choice of scaling for $\eta_A$ we choose $\eta_C = \gamma/\mu$ so that $\tilde{C}_* \sim 1$ as desired. A natural choice of timescale is 
$\eta_t = 1/\mu$. Then letting $\epsilon = \frac{\mu\eta_t\eta_C}{\eta_A} \sim 10^{-2}$ this implies the nondimensional form of these equations are  

\begin{align}
    \frac{\partial A}{\partial t} &= \epsilon C + \rho_0 A(1 - A), \\
    \frac{\partial C}{\partial t} &= \frac{D}{\mu\eta_x^2}\frac{\partial^2 C}{\partial x^2} - C + \frac{A^2}{\beta^2 + A^2}.
\end{align}

\noindent where the $\sim$ have been dropped and parameters have been amalgamated for clarity. Note that now we have all parameters are $\mathcal{O}(1)$ except $\epsilon \sim 10^{-2}$. 
Furthermore we have that $\epsilon C_* \ll 1$ and so the fixed points of $A$ must approximately be 

\begin{equation}
    A_* = \frac{1 \pm \sqrt{1 + 4\rho_0^{-1}\epsilon C^*}}{2} \approx 1 \pm 1 + \mathcal{O}(\epsilon).
\end{equation}

So for the fixed point away from the origin (e.g.\! that corresponding to a fully forested domain) we obtain that $\epsilon C_*$ has a negligible impact on the location 
of the fixed point as required.

\par
It remains to pick a spatial length scale such that $\hat{D} = D/(\eta_x^2\mu)$ is of an appropriate size. Without detailed spread data it is 
difficult to estimate $D$, but fortunately it is easier to determine the correct choice of $\eta_x$ so that 
the non-dimensional group $\hat{D}$ is $\mathcal{O}(1)$. Based on results given in \cite{mythesis} we expect that spread rates are bounded above by 
$\sqrt{\hat{D}}\rho_0$ and in \cite{caplat2012seed} we are given that spread rates are approximately 80m/year (at least in the New Zealand context), 
so for $\hat{D},\rho_0 \sim 1$ we require that  $\sqrt{\hat{D}}\rho_0 \eta_x/\eta_t \approx 1 \eta_x/\eta_t \approx 80$. Thus we have that $\eta_x \sim 10^3$ 
so our scale for $x$ is approximately one of kilometres. Consequently (dropping the carat on $D$) we 
obtain the final nondimensional model. 

\begin{align}
    \frac{\partial A}{\partial t} &= \epsilon C + \rho_0 A(1 - A), \\
    \frac{\partial C}{\partial t} &= D\frac{\partial^2 C}{\partial x^2} - C + \frac{A^2}{\beta^2 + A^2}.
\end{align}

\section{Numerical Results}\label{sec:num_res}

One natural question is whether the (nondimensional) model produces results which are qualitatively similar to the dynamics we expect for this scenario. To explore the qualitative dynamics of our model we implemented a 
second-order in space first-order in time finite differences scheme, using the 
python programming language and the \verb|scipy|, \verb|numpy| and \verb|matplotlib| libraries. We used the following parameter values 

\begin{table}[H]
    \centering
\begin{tabular}{|p{0.1\textwidth}|p{0.07\textwidth}|p{0.07\textwidth}|p{0.07\textwidth}|p{0.07\textwidth}|p{0.07\textwidth}|p{0.07\textwidth}|p{0.07\textwidth}|p{0.07\textwidth}|}
    \hline
    Name & $\epsilon$ & $\rho_0$ & $D$ &$\beta$ & $L$ & $n$ & $\Delta_x$ & $\Delta_t$\\
    \hline 
    Values & $10^{-2}$ & 1 & 1 & 1 & 5 & 50 & $10^{-1}$ & $10^{-3}$ \\ 
    \hline
\end{tabular}
\caption{Parameter Values Chosen.}
\end{table}

\par
\begin{figure}[H]
    \centering
    \includegraphics{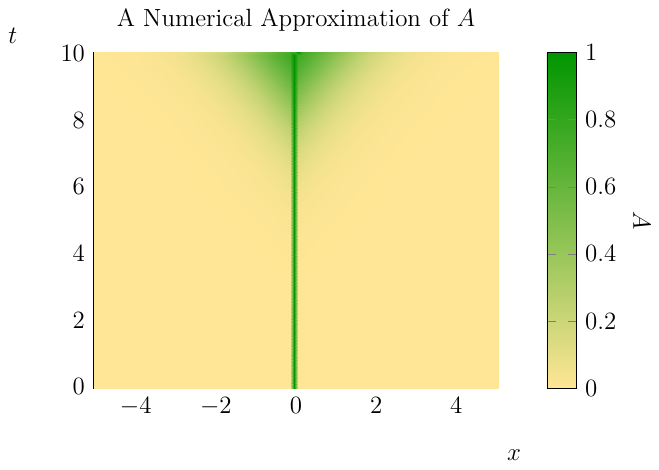}
    \caption{Simulation of an invasion from $t = 0$ to $t = 10$, tracking adult density only.}
    \label{fig:num_adult_sim}
\end{figure}

Solutions were plotted 
between $t = 0$ and $t = 10$ to observe the dynamics of the system for $t \sim 1$. Since $t = 10$ corresponds to one hundred years after the 
initial invasion, it is biologically reasonable that we observe adult trees expanding to cover the domain (which is approximately ten kilometres wide) towards the 
end of our simulation (see \autoref{fig:num_adult_sim}).

\par 
Notice that this numerical evidence suggests that our simple model produces a simulation of the 
dynamics that appears (at least phenomenologically) reasonable. In particular, we relatively observe 
little spread initially, followed by a rapid `burst' of growth. This matches the behaviours 
observed in \cite{caplat2012seed} although it should be noted that 
\citeauthor{caplat2012seed} attribute most of this delay to spatial variation in wind patterns. 
Furthermore, for intermediate to large times we see that the solutions tends toward a travelling 
wave type structure, suggesting an asymptotically constant spread rate (like those sought in 
\cite{neubert2000demography}). 

\section{Semi-analytical approaches}
As a first approach to asymptotic analysis of this problem, we can attempt a regular perturbation analysis of our nondimensional system 

\begin{align}
    \frac{\partial A}{\partial t} &= \epsilon C + \rho_0 A(1 - A), \\
    \frac{\partial C}{\partial t} &= D\frac{\partial^2 C}{\partial x^2} - C + \frac{A^2}{\beta^2 + A^2}.
\end{align}

\noindent With the boundary conditions $C_x(-L,t) = C_x(L,t) = 0$ and initial condition $C(x,0) = 0$, $A(x,0) = h_n(x)$, $n > 1$, where we have 

\begin{equation}
    h_n(x) = \begin{cases}
        1 & x \in [-L/n,L/n] \\
        0 & \text{otherwise}
    \end{cases}
\end{equation} 

Given the presence of the small parameter $\epsilon$ we consider the ansatz;

\begin{align}
    A &= A_0 + \epsilon A_1 + \epsilon^2 A_2 + \dots \\
    C &= C_0 + \epsilon C_1 + \epsilon^2 C_2 + \dots
\end{align}

\noindent Since we desire that our expression is asymptotically valid for small values of $\epsilon$ we assume 
that $C_{i,x}(-L,t) = C_{i,x}(L,t) = 0$ for all $i$ and the 
initial condition that $C_i(x,0) = 0$ for all $i$ and $A_0(x,0) = h_n(x)$, $A_i(x,0) = 0$ for all $i > 0$. The leading terms in this series are now

\begin{align}
    \frac{\partial A_{0}}{\partial t} &= \rho_0 A_0(1 -  A_0), \\
    \frac{\partial C_0}{\partial t} &= D\frac{\partial^2 C_0}{\partial x^2} - C_0 +  \frac{A_0^2}{\beta^2 + A_0^2}.
\end{align}

It is now possible to solve these equations fairly directly. Considering first our equation for $A_0$ this has the following solution

\begin{align}
    A_0(x,t) = 
    \begin{cases}
        1 & x \in [-L/n,L/n], \\
        0 & \text{otherwise}.
    \end{cases}
\end{align}

\noindent 
With this solution in hand it is possible to solve for $C_0$ and obtain 

\begin{multline}
    C_0(x,t) = \frac{1}{n(\beta^2 + 1)}(1 - e^{-t}) + \\
    \sum_{k=1}^{\infty} \left(\frac{2 L^2}{\pi k(\beta^2 + 1)(D\pi^2k^2 + L^2)}\sin\left(\frac{\pi k}{n} \right)\right)\\
    \left(1 - e^{-\left(\frac{D\pi^2k^2}{L^2} + 1\right)t}\right)\cos\left(\frac{\pi kx}{L}\right).
\end{multline} \label{eq:C_0_expression}

\noindent Considering now $A_1$, we see that away from the support of the initial condition (e.g.\! for $x \notin [-L/n,L/n]$) the problem reduces to 

\begin{equation}
    \frac{\partial A_1}{\partial t} = C_0 + \rho_0A_1.
\end{equation}

\noindent Since $C_0$ is a bounded and strictly positive function it follows that for $t \gg 0$ the solution to the first-order term will be dominated by a term of the 
form $Fe^{\rho_0 t}$, $F \in \mathbb{R}$ and so for $t > \ln(\epsilon^{-1})/\rho_0$ this regular approximation to the solution will break down. This suggests that a regular perturbation 
approach is not suitable for this problem. 

\subsection{Singular Perturbation Approaches}
The failure of the regular perturbation approach above appears to occur because the neglect of the order $\epsilon$ terms is not appropriate near the 
origin, where $A(x,t) \ll 1$ for $x \notin [-L/n,L/n]$. Consequently, we now consider inner and outer expansions and attempt to match these for some intermediate 
$t$. In the inner region (e.g.\! for $t$ near zero) we expect that $A = 1 + \mathcal{O}(\epsilon)$ for $x \in [-L/n,L/n]$ and that $A \sim \epsilon$ 
elsewhere. Thus for $x \notin [-L/n,L/n]$ we rescale $\hat{A} = \epsilon^{-1}A$ to obtain 

\begin{equation}
    \epsilon \frac{\partial \hat{A}}{\partial t} = \epsilon C + \epsilon\rho_0 \hat{A} + \mathcal{O}(\epsilon^2).
\end{equation}

\noindent Given that we have already assumed that the order one expression for $A = h_n(x)$ at leading order we have 

\begin{equation}
    \frac{\partial \hat{A}}{\partial t} = C_0 + \rho_0 \hat{A},
\end{equation}

\noindent where $C_0$ is equal to the expression given in \autoref{eq:C_0_expression} and $A(x,0) = 0$. This first-order equation can be solved directly, so that 

\begin{align}
    A_{\text{inner}}(x,t) &= h_n(x) + (1 - h_n(x))\epsilon \nonumber\\
    &\sum_{k=0}^{\infty}\alpha_k\left(\frac{e^{\rho_0 t} - 1}{\rho_0} - \frac{e^{\rho_0 t} - e^{-(D\omega_k^2 + 1)t}}{D\omega_k^2 + 1 + \rho_0}\right)\cos(\omega_k x), \\
    \alpha_k &= \begin{cases}
        \frac{1}{n(\beta^2 + 1)} & k = 0, \\
        \frac{2}{\pi k(\beta^2 + 1)(D \omega_k^2 + 1)} & k > 0,
    \end{cases} \\
    \omega_k &= \frac{\pi k}{L}.
\end{align}

\noindent Since this expression for $A$ is increasing exponentially in time, it cannot be valid everywhere. In particular, we expect that if $t \sim \ln(\epsilon^{-1})/\rho_0$ 
this approximation will break down, as our assumption that $\rho_0 A \sim \epsilon C_0$ is clearly false. 

\par 
Therefore, for $t \gg \epsilon$ we might expect $A$ to be $\mathcal{O}(1)$ and thus the leading order dynamics of $A$ (for $x \notin [-L/n,L/n]$) are given by 

\begin{equation}
    \frac{\partial A}{\partial t} = \rho_0 A (1 - A).
\end{equation}

\noindent Thus we have 

\begin{equation}
    A_{\text{outer}}(x,t) = \frac{Ge^{\rho_0 t}}{1 + Ge^{\rho_0 t}}, \qquad G \in \mathbb{R}^+.
\end{equation}

We expect that these two approximations to the solution should be approximately equal at the 
point of transition between the inner and outer regions and thus should coincide for some thin region about $t_* = \ln(\epsilon^{-1})/\rho_0$. If we consider a 
band of thickness $\epsilon$ about $t_*$ and $x \notin [-L/n,L/n]$ then one can write the inner expansion as follows

\begin{align}
    A_{\text{inner}}(x,t) &= \epsilon \sum_{k=0}^{\infty}\alpha_k\Biggl(\frac{e^{\rho_0 (t - t_*)}e^{\rho_0 t_*} - 1}{\rho_0} \nonumber \\
    &- \frac{e^{\rho_0 (t - t_*)}e^{\rho_0 t_*} - e^{-(D\omega_k^2 + 1)t}}{D\omega_k^2 + 1 + \rho_0}\Biggr)\cos(\omega_k x), \\
    &= \sum_{k=0}^{\infty}\alpha_k\left(\frac{1}{\rho_0} - \frac{1}{D\omega_k^2 + 1 + \rho_0}\right)\cos(\omega_k x)e^{\rho_0 (t - t_*)} + \mathcal{O}(\epsilon).
\end{align}

Similarly, if we let $G \sim \epsilon$ then one can write 

\begin{align}
    A_{\text{outer}}(x,t) &= \frac{Ge^{\rho_0 (t - t_*)}e^{{\rho_0 t_*}}}{1 + Ge^{\rho_0 (t - t_*)}e^{\rho_0 t_*}}, \\
    &= \frac{\epsilon^{-1} Ge^{\rho_0 ( t- t_*)}}{1 + \epsilon^{-1} Ge^{\rho_0 ( t- t_*)}}.
\end{align}

\noindent Considering now an expansion of $1/(1 + \epsilon^{-1}Ge^{\rho_0(t - t_*)})$ as a Taylor series about $t_*$ one obtains 

\begin{equation}
    \frac{1}{1 + \epsilon^{-1}Ge^{\rho_0 ( t- t_*)}} = \frac{1}{1 + \epsilon^{-1}G} - \frac{\rho_0 \epsilon^{-1}G}{(1 + \epsilon^{-1}G)^2}(t - t_*) + \mathcal{O}((t - t_*)^2).
\end{equation}

\noindent Furthermore, since $t - t_* \sim \epsilon$ it follows that $e^{\rho_0 (t - t_*)} \sim 1$ and thus 

\begin{equation}
    A_{\text{outer}} = \frac{\epsilon^{-1} G}{ 1 + \epsilon^{-1}G}e^{\rho_0(t - t_*)} + \mathcal{O}(\epsilon).
\end{equation}

\par
Matching at leading order one obtains 

\begin{align}
    \frac{\epsilon^{-1} G}{1 + \epsilon^{-1}G} &= \sum_{k=0}^{\infty}\alpha_k\left(\frac{1}{\rho_0} - \frac{1}{D\omega_k^2 + 1 + \rho_0}\right)\cos(\omega_k x), \\
    \implies G &= \epsilon \frac{\nu}{1 - \nu}, \\
    \nu &= \sum_{k=0}^{\infty}\alpha_k\left(\frac{1}{\rho_0} - \frac{1}{D\omega_k^2 + 1 + \rho_0}\right)\cos(\omega_k x).
\end{align}

\noindent Given that we have only one free parameter, we cannot proceed further with matching. Thus we have obtained a pair of matched solutions, each valid 
in a part of the domain. It remains to obtain a uniformly valid expansion. For `thin' initial conditions, $\nu$ is small (the size of the first Fourier coefficient is 
$\mathcal{O}(n^{-1})$) and from numerical evidence that we expect that for $n \sim \epsilon^{-1}$ and as $t$ approaches zero one can write 

\begin{equation}
    A_{\text{outer}} = \epsilon\nu e^{\rho_0 t} + \mathcal{O}(\epsilon^2).
\end{equation}

\noindent Subtracting the common part in this region of the domain then leads to the following expression for $A$:

\begin{multline}
    A = h_n(x) + \\(1 - h_n(x))\left(\frac{Ge^{\rho_0 t}}{1 + Ge^{\rho_0 t}} - \epsilon \sum_{k=0}^{\infty}\alpha_k\left(\frac{1}{\rho_0} - \frac{e^{-(D\omega_k^2 + 1)t}}{D\omega_k^2 + 1 + \rho_0}\right)\cos(\omega_k x)\right) + \mathcal{O}(\epsilon^2).
\end{multline}

\noindent Numerical evidence suggests that this approximation is $\mathcal{O}(\epsilon)$ for $t > \ln(\epsilon^{-1})/\rho_0$ and $\mathcal{O}(\epsilon^2)$ 
although relative error increases rapidly as $t$ approaches zero and the magnitude of solutions decreases sharply (see the following section). Thus we expect an overall first-order 
accurate approximation of $A$ to be 

\begin{equation}
    A(x,t) = \begin{cases}
        1 & x \in [-L/n,L/n], \\
        \frac{Ge^{\rho_0 t}}{1 + Ge^{\rho_0 t}} - \epsilon \sum_{k=0}^{\infty}\alpha_k\left(\frac{1}{\rho_0} - \frac{e^{-(D\omega_k^2 + 1)(t - t_*)}e^{-\rho_0 t_*}}{D\omega_k^2 + 1 + \rho_0}\right)\cos(\omega_k x) & \text{otherwise}.
    \end{cases}
\end{equation}

\subsection{Comparison of Numerical and Approximate Results}

To compare these approximate results obtained with asymptotic analysis to numerical evidence, return to the numerical results discussed in \autoref{sec:num_res}. For comparison purposes, \verb|numpy|'s \verb|fft| 
library was used to obtain accurate estimates of the Fourier coefficients $\alpha_k$. Numerical evidence suggested that using \verb|fft| (which produces an approximation to the Fourier coefficients)
produced better results than resolving a finite number of the Fourier coefficients discussed in previous sections. We suspect that this is due to the 
discontinuous nature of $A_0$ leading to slow convergence of the Fourier coefficients computed above, although we do not prove this. With the 
first fifty coefficients thus calculated it was possible to obtain a seemingly reasonable 
approximation to the solution.

\begin{figure}[H]
    \centering
    \includegraphics{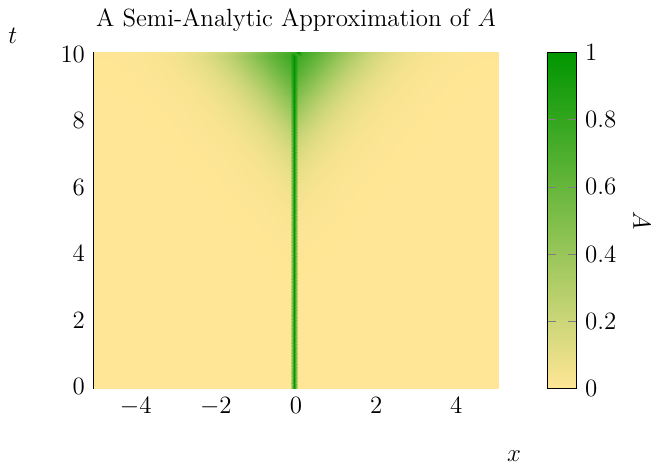}
    \caption{An approximate solution of an invasion from $t = 0$ to $t = 10$, tracking adult density only.}
    \label{fig:approx_adult_sim}
\end{figure}

\par 
We can now see that the semi-analytical and numerical solutions give qualitatively similar behaviour, 
which gives some confidence that these approximate solutions are accurate. To make this qualitative 
similarity more precise, we consider first the absolute difference and then the relative difference 
between these two solutions. 

\begin{figure}[H]
    \centering
    \includegraphics{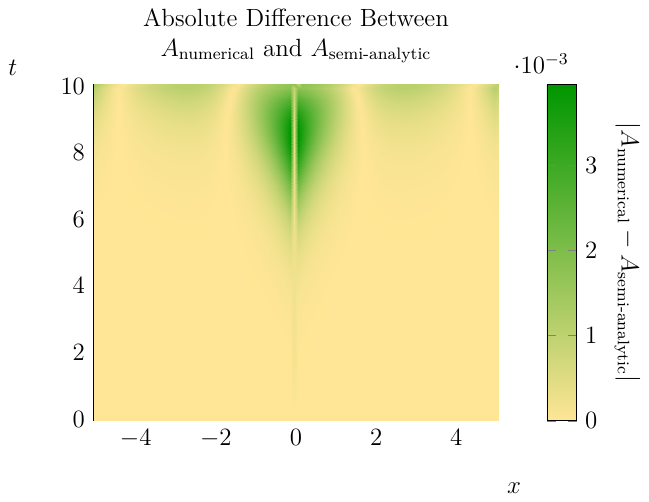}
    \caption{Absolute difference between the two solutions from $t = 0$ to $t = 10$, tracking adult density only.}
    \label{fig:abs_diff}
\end{figure}

\noindent 
Notice that the absolute difference is $\mathcal{O}(\epsilon)$, as we predicted in the 
previous section. We can now consider the relative error, which in this case we will 
define to be $R = \ln(|(A_{\text{numerical}} - A_{\text{semi-analytic}})/A_{\text{semi-analytic}}|)$. Note that 
we are plotting the log of this quantity to give better insight into both large and small error values, omitting the two first two data points where $|A_{\text{numerical}} - A_{\text{semi-analytic}}| \ll \epsilon$ 
and thus numerical approximations of the natural logarithm produced division by zero errors. Fortunately 
it is easy to see that errors remain relatively small everywhere, except for a few isolated 
points. It seems likely that the large errors at these points were caused by resolving only a 
finite number of Fourier coefficients.

\begin{figure}[H]
    \centering
    \includegraphics{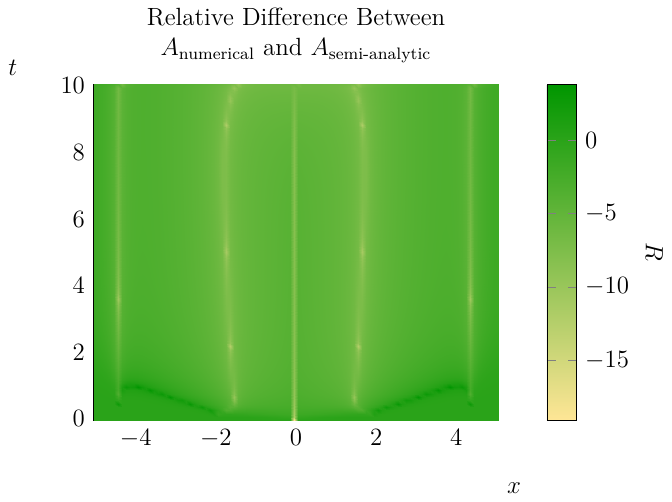}
    \caption{Relative difference between the semi-analytical solution and the numerical solution from $t = 0.002$ to $t = 10$, tracking adult density only.}
    \label{fig:relative_diff}
\end{figure}

Given that absolute error remains $\mathcal{O}(\epsilon)$ everywhere and relative error is $\mathcal{O}(1)$ or less 
except at isolated points, our semi-analytical and numerical approaches are broadly aligned. This 
gives confidence in our predictions, as two separate techniques give similar results. Therefore we can feel confident in saying that the solution exhibits distinctly 
different behaviour for $t \ll t_* = \ln(\epsilon^{-1})/\rho_0$ and for $t \gg t_*$. For an $\mathcal{O}(1)$ period after $t =0$ the solution exhibits 
almost no change, before rapidly evolving towards a travelling wave type structure after $t_*$.  

\section{Discussion} 
\subsection{Limitations of Previous Models}\label{subsec:lim_past_models}
In \cite{caplat2012seed} and \cite{buckley2005slowing} two models for wilding pine spread are advanced. Both are discrete in time but continuous in space and seek 
to model spread with a matrix integrodifference equation. However rather than simulating a nonlinear model (which would be difficult 
to parametrise but would provide a good description of the spread process) both papers advance a linear model for spread. While it is possible to show that for a 
given linear model the asymptotic spread speeds of this model are equal to the asymptotic spread speeds of a suitable nonlinear model \cite{weinberger2002analysis}, the asymptotic spread speeds of 
the linear models considered in both papers are not known. Both \cite{caplat2012seed} and \cite{buckley2005slowing} rely on results presented in \cite{neubert2000demography}, 
but this work only gives an upper bound on spread speeds and conjectures that this is also the asymptotic spread speed of these linear models.

\par 
Setting aside these theoretical concerns, there are a number of practical concerns with these models (in particular with \cite{caplat2012seed}). In particular, 
both models assume that all seed dispersed from trees in the landscape is dispersed along a single line. This likely significantly overestimates the true amount of 
seeds dispersed in any single direction, as observed wind data at invasion sites (see \cite{caplat2012seed}) suggests that there are multiple different wind 
directions which may contribute to spread (see also the spread patterns shown in Figure 1 of \cite{caplat2012seed} and Figure 1 of \cite{sprague2021density}). 
In some cases, parameter values reported contradict biological intuition. For example the average juvenile retention rate given in \cite{caplat2012seed} is 
$r_J = 0.78$ which suggests that in a mature forest $22\%$ of trees became reproducing at age two, contradicting existing data \cite{coutts2012reproductive}. 
Further issues include numerical instability, sensitivity to distributional assumptions, and difficulties interpreting outputs, given that the model is parametrised 
in terms of discrete counts of trees in locations but returns fractional distributions of trees across a continuous landscape. 

\subsection{PDE Based Models for Wilding Pine Spread}
As well as reviewing past models for wilding pine spread, we present a simple PDE model for wilding pine spread. In this model we move from tracking individual trees to considering the evolution of quantities 
of biomass across the landscape. In particular, we track the quantity of `adult' biomass (e.g.\! that from reproducing trees) and `child' or potential adult biomass 
in each location. The evolution of these quantities is then described by the following PDE (after suitable nondimensionalisation),

\begin{align}
    \frac{\partial A}{\partial t} &= \epsilon C + \rho_0 A(1 -A), \\
    \frac{\partial C}{\partial t} &= \nabla^2D C - C + \frac{\gamma A^2}{\beta^2 + A^2},
\end{align}

\noindent where $\epsilon \ll 1$ and all other parameters are $\mathcal{O}(1)$. We consider this model in both one-dimensional and two-dimensional domains, but 
in all cases we impose no-flow boundary conditions to approximate the case where a landscape is bounded by impassible terrain (e.g.\! valley walls). In both one and two 
dimensional cases we impose initial conditions which require $A$ to be $\mathcal{O}(1)$ within a small distance of $x=0$ and zero elsewhere ($C$ is set to 
zero everywhere). Because only limited data was available, we set all parameters to one except $\epsilon$, which was taken to be $\epsilon = 10^{-2}$. Numerical 
simulations of this model revealed a qualitative picture of invasions, where little spatial spread is observed until some time $t_* \gg 0$ after which the system 
rapidly evolves towards a travelling wave solution of the form $A(x,t) = f(|-x+st|)$, $s \in \mathbb{R}^+$.

\par 
To gain a deeper understanding of the dynamics of the system, we attempted to use a singular perturbation approach to approximate the solution to the PDE. Outside 
of a thin band around the origin $A(x,0)$ is zero, so the problem was rescaled so that $A \sim \epsilon$ outside this region. This leads to a solution for $A$ which is close to $\epsilon C e^{\rho_0 t}$ in character and valid until $ t \sim \ln(\epsilon^{-1})/\rho_0$. Outside of this region, the leading order solution to this problem $A(x,t) = G(x)e^{\rho_0t}/(1 + G(x)e^{\rho_0t})$ 
is valid. Solutions were thus matched in a band of width $\epsilon$ about $t_* = \ln(\epsilon^{-1})/\rho_0$ to obtain a globally valid approximation to the solution. This solution and 
the numerical solution had an absolute difference of $\mathcal{O}(\epsilon)$ across the domain and between $t=0$ and $t=10$.

\par 
This approximate solution confirms our qualitative impressions that the regime that governs the behaviour of solutions for $t \ll t_*$ 
is very different to the regime that governs for $t \gg t_*$. In particular, for small $t$ (although note that this region is still $\mathcal{O}(1)$), 
the solution is $\mathcal{O}(\epsilon)$ (hence the lack of change observed numerically). Furthermore, at least for some $t \gg t_*$ the solution is determined 
by the solution to $C$ for small times. Structurally we expect that, away from the origin, for small $t$ the decay in $C$ as $x \rightarrow \pm \infty$ is approximately 
exponential which leads to the travelling wave behaviour we see in the numerical solutions (but we do not prove this). 

\section{Directions for Future Work}
\subsection{Discrete Time Models for Spread}
Because of these issues discussed in \autoref{subsec:lim_past_models}, both \cite{caplat2012seed} and \cite{buckley2005slowing} are unlikely to provide an accurate prediction of spatial spread rates and should 
not be used to generate management recommendations. However, discrete-time-continuous-space or discrete-time-discrete-space models are not inherently flawed 
as modelling tools for wilding pine spread. The models presented in \cite{caplat2012modeling} or \cite{davis2019simulation} are discrete-time-discrete-space 
models for wilding pine spread which, if modified and parametrised for the New Zealand context, could be a suitable approach to modelling pine spread. However, 
such models are dependent on an appropriate spatial discretisation ensuring that the dynamics of spread are accurately captured. While previous models discretise 
space by dividing the domain into square or rectangular cells, further research should consider whether alternative geometries may better capture the dynamics 
of spread. Network models may also be an appropriate tool, especially for simulating macro-level spread between different sites.

\par 
Alternatively, discrete-time-continuous-space models could obviate concerns about the spatial discretisation of the domain. Either a fully-nonlinear integrodifference 
model or an agent-based approach could provide valuable insights. In both cases, since invasions often lead to dense populations of trees in particular 
locations, models will need to account for shading effects or local resource constraints. Such shading effects may be computationally intensive for agent-based 
models, although agent-based approaches may still be valuable for understanding early invasion dynamics.

\subsection{Extending Our Model}
While we have conducted some initial analysis of the model we present in this paper, there remain a number of further questions which future work could explore. Firstly, we have 
not attempted to prove existence or uniqueness of solutions. Determining existence and uniqueness in suitable spaces could be a useful theoretical direction for 
future work to explore. Alternatively, while the approach we employed did give insight into why the solution exhibited a long quiescent period followed by 
rapid growth, we did not determine why the solution appeared to converge towards a particular travelling wave solution. Future work could give greater insight into the 
reasons why particular travelling wave speeds were selected by the model.

\par 
This travelling wave speed may be a particularly fruitful area for further research, as it is equivalent to the spread rate computed in \cite{caplat2012seed} and \cite{buckley2005slowing}. 
In forthcoming work, we advance a computational approach for estimating the spread rate. Given this computational approach, even purely numerical attempts 
to investiage the sensitivity of the spread rate to variation in parameters could potentially be very fruitful. Such work would also be complemented by fitting 
the model to data and estimating more accurately the value of parameters in an at-risk environment.

\par
Many of these are easy to interpret and estimate from data, while 
some (e.g.\! $D$, $\kappa$, $\gamma$ and $\beta$) may lack obvious interpretations or be harder to identify. Such parameters will need to be fitted from data (possibly like that reported in \cite{bi2010additive} or \cite{madgwick1994pinus}) 
or otherwise inferred, so understanding how statistical error in parameter estimation might impact the accuracy of predicted spread rates may be valuable. 

\par 
Beyond further analysis or further application of the model presented here, further research could also explore structural changes to the model. For example, our model 
features a very simple stage structure which combines all non-reproducing biomass into a single category. This approach vastly simplifies analysis of the model, 
but is necessarily reductive. A more complex approach with a more sophisticated stage structure may give greater insights, although it should be noted that it 
may be difficult to include more classes of non-reproducing trees without considering multiple timescales, as the timescale at which adult biomass evolves is 
necessarily very different to the timescale which governs seed dispersal.

\par
Because our PDE model allows for explicit simulation of solutions, it is relatively straightforward to incorporate spatial heterogeneity in parameters. In 
forthcoming work we will demonstrate that spatial variation can have a significant impact on solutions, depending on the parameter and the magnitude of variation 
considered. Future work could also explore the impact of different management strategies on spread (as in \cite{caplat2014cross}) 
or environmental phenomena that also evolves over time (e.g.\! coinvasion by herbivores). These time-varying phenomena could potentially have significant impacts on 
the dynamics of solutions and insights from such models may have particularly valuable management implications so future work in this area could be productive.

\subsection{Preferred Directions of Spread}\label{subsec:pref_direc_spd}
As well as altering stage structure or incorporating spatial variation, further work could explore alternate models of spread. Our current simple model assumes dispersal in the non-reproducing category 
is diffusive, implying spread occurs by an isotropic process. The implied dispersal kernel was a Laplace distribution, suggesting an exponential decay away 
from the source tree. While this kernel is leptokurtic (e.g.\! heavy tailed) as is generally considered desirable for modelling windborne spread (see \cite{katul2005mechanistic}) 
the isotropic assumption is known to be false in observed invasions (see \cite{sprague2021density} or \cite{caplat2012seed}). A more sophisticated approach 
could incorporate a preferred direction of spread, most simply by including an advective term. Without the assumption of isotropic spread (although we do assume that diffusion remains isotropic), this model is more naturally 
suited to a 2D domain, leading to the following equation  

\begin{align}
    \frac{\partial A}{\partial t} &= \epsilon C + \rho_0 A(1 - A), \\
    \frac{\partial C}{\partial t} &= D\nabla^2 C + \nu \cdot \nabla C - C + \frac{A^2}{\beta^2 + A^2}.
\end{align}

\begin{figure}[H]
    \centering
    \includegraphics{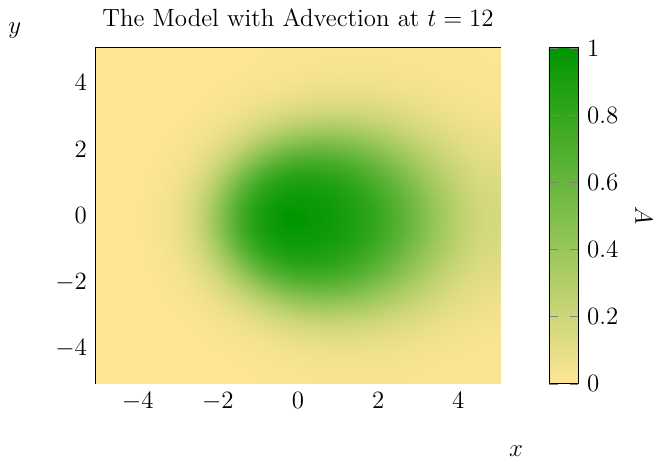}
    \caption{A snapshot (at $t=12$) of mixed advective-diffusive spread from a point source on a 2D domain.} 
    \label{fig:advective_example}
\end{figure}

\noindent 
It is relatively straightforward to model this scenario with 
a upwind finite-differences method, either in one dimension or as a 2D model (see \autoref{fig:advective_example} for a 2D example with a $\mathcal{O}(1)$ advective 
term).

\par
It also seems likely that the approach we present in this paper 
would also produce a good semi-analytic approximation of the dynamics of the modified equation, although we have not confirmed this. More complex models (e.g.\! those incorporating fractional diffusion) could also be considered to obtain dispersal kernels that closely 
match those derived from empirical data \cite{baeumerperscomm}.

\section{Conclusion}
Invasive exotic conifers are a significant threat to ecologically important landscapes in New Zealand or elsewhere. In order to combat this threat, conservation managers need 
information on predicted future spread and the factors that mitigate or accelerate invasions. Mathematical modelling approaches have often been used to answer 
these questions, although past models have a number of shortcomings that reduce the reliability of their predictions. 

\par
However, since parameter estimates 
which are used as inputs to this model do not reflect biological reality and the theoretical underpinnings of these IDM models are weak, we propose an alternate model for wilding pine spread. Our PDE approach is straightforward 
to simulate numerically and benefits from the rich variety of analysis techniques already developed for PDEs. Using our model, and parameters which are accurate only 
to the nearest order of magnitude, we obtain estimates for asymptotic spread rates which are surprisingly close to the average spread rates reported in real invasions (note that while our choice 
of $D$ was selected so that estimated asymptotic spread rates were of a similar order of magnitude to those observed, it is still surprising that our estimates are within 10m/year of the 
true observed average spread rates). 
Furthermore we apply singular perturbation techniques to develop an accurate approximation to the behaviour of our model, revealing that our model has two main 
regimes and that solutions transition from one kind of behaviour to another at some critical time (matching observed behaviour in some field cites, see \cite{caplat2012seed}). 

\par 
Future work could look to answer questions of existence and uniqueness of solutions to the PDE model itself, or to fit the model to real world data. While it 
is clear that travelling wave solutions to the model are possible, understanding why certain initial conditions lead to certain wave speeds would improve 
our understanding of the model and may have management implications. Better awareness of the impact of spatial heterogeneity in different parameters, as well as 
a clearer picture of which parameters can be expected to vary significantly in space, may also improve our ability to better fit the model to data.

\par 
Because of the risks that uncontrolled wilding pine spread poses to New Zealand landscapes, insights arising out of a deeper understanding of the dynamics of spread of invasive pines is 
particularly valuable to conservation practitioners. It is hoped that this thesis will advance our understanding of these dynamics by uncovering flaws with 
previous models, advancing an alternative approach to model invasions, and by suggesting avenues for further research that may be fruitful.

\bibliography{Thesis}

\end{document}